\documentclass{article}
\usepackage{geometry} 
\usepackage{xcolor}
\usepackage{graphicx}
\usepackage[utf8]{inputenc}
\usepackage{amsmath}
\usepackage{float}
\usepackage[
    backend=biber,
    style=numeric,
  ]{biblatex}
\title{How norms shape the evolution of prosocial behavior \\ \hrulefill \\ \large 
Compassion, Universalizability, Reciprocity, Equity: A C.U.R.E for social dilemmas} 
\author{Brian Mintz, Feng Fu} 

\begin{document}

\maketitle

\begin{abstract}
   How cooperation evolves and particularly maintains at a large scale remains an open problem for improving humanity across domains ranging from climate change to pandemic response. To shed light on how behavioral norms can resolve the social dilemma of cooperation, here we present a formal mathematical model of individuals' decision making under general social norms,  encompassing a variety of concerns and motivations an individual may have beyond simply maximizing their own payoffs. Using the canonical game of the Prisoner's Dilemma, we compare four different norms: compassion, universalizability, reciprocity, and equity, to determine which social forces can facilitate the evolution of cooperation, if any. We analyze our model through a variety of limiting cases, including weak selection, low mutation, and large population sizes. This is complemented by computer simulations of population dynamics via a Fisher process, which confirm our theoretical results. We find that the first two norms lead to the emergence of cooperation in a wide range of games, but the latter two do not on their own. Due to its generality, our framework can be used to investigate many more norms, as well as how norms themselves emerge and evolve. Our work complements recent work on fair-minded learning dynamics and provides a useful bottom-up perspective into understanding the impact of top-down social norms on collective cooperative intelligence.
\end{abstract}


\section{Introduction}
Through millennia, evolution has produced incredibly sophisticated mechanisms by which organisms manage to survive and reproduce~\cite{axelrod1981evolution}. It is often difficult to explain how these arise through natural selection, as any sufficiently complex trait would likely result from a series of mutations that are likely neutral, or possibly deleterious on their own~\cite{perc2017statistical}. One prominent example of this is the intricate social systems seen throughout the natural world, from wolf packs to insect colonies. Vampire bats are known to cooperate in numerous ways, from sharing food to even their own blood \cite{wilkinson2016non}. Stickleback fish have also been observed cooperating to handle predators \cite{milinski1987tit}. It remains a longstanding question how such complexity could form emerge though the simple process of mutation and selection, especially since cooperation often entails some degree of cost that may not be compensated for \cite{nowak2006five}. The canonical example of the tension between acting in a mutually beneficial way or the more compelling, selfish alternative is the Prisoner's Dilemma. Introduced in 1950 by Merrill Flood and Melvin Dresher, but named by Albert Tucker, this game is a concrete version of this problem, assigning payoffs based on the which actions, selfish or cooperative, each individual chooses. Specifically, the below matrix gives the payoff received by a player given their action, in the row, and the other player's action, in the column. The possible actions are to cooperate or defect (C or D), yielding one of four possible payoffs $S<P<R<T$~\cite{doebeli2005models}:
\begin{equation}
        \begin{tabular}{c|cc}
            & C & D\\
            \hline
            C & R & S \\
            D & T & P
        \end{tabular}
        \label{pdmatrix}
 \end{equation}
 
 Note these inequalities mean that regardless of the other player's choice, it is always optimal to defect, so individuals who are acting in self interest will choose to defect, leading to a worse outcome than mutual cooperation. Indeed, this is what rationality would suggest is the appropriate strategy. However, behavioral studies in humans show that people don't always follow the rational, albeit selfish strategy. Psychologists have proposed numerous explanations why people don't behavior rationally, including intelligence or personality, or that individuals adhere to some social norm \cite{jones2008smarter, boone1999impact, fehr2002strong}. In this work, we ask whether a mathematical model of four norms can promote cooperation. Specifically, we investigate compassion, universalizability, reciprocity, and equity. By prescribing a degree to which individuals follow or care about a given norm, we can model decision making that considers both individual payoff and broader social norms.

The four norms we model in this work have long histories in theories of morality. Compassion, also known as empathy, means caring about others rather than just oneself, and is one of the main factors theorized to influence cooperative behavior \cite{chierchia2017neuroscience, batson1999empathy}. Universalizability is a concept introduced by the philosopher Immanuel Kant in the 18th century that claims an action's morality is determined by its effects when adopted by everyone. Immoral actions, in this framework, are those that are detrimental when they become universal \cite{narveson1985and, reinikainen2005golden}. While this may seem like too sophisticated to apply to simpler organisms, it also makes evolutionary sense, as any successful trait should remain beneficial should it spread throughout a population. Further, it also describes the consequence of kin-interactions, as they likely share similar behavior. This norm is a bit stricter than compassion, as it additionally forbids actions like littering or free-riding, which do not harm any individual in particular, but only become truly problematic when everyone does them. Reciprocity is the exchange of beneficial actions~\cite{wolf2011coevolution}, and has been found to be a key aspect in many experiments with social dilemmas, as well as an integral part of many cultures throughout the world \cite{gachter2009reciprocity, clark2001sequential, komorita1991reciprocity}. Lastly, equity, or fairness, is a desire for unbiased treatment~\cite{chen2023outlearning}, and is widely valued across different communities~\cite{nowak2000fairness}. Interestingly, even animals have been observed to care about fairness. Frans De Waals conducted a fascinating experiment which showed capuchin monkeys frustration with unequal rewards for the same task \cite{brosnan2003monkeys, brosnan2012fairness}. This suggests that this factor may be a driving force in the evolution of cooperation, since it appears throughout the animal kingdom. 


Apart from these considerations, evolutionary biologists have theorized several other mechanisms that can promote cooperation~\cite{nowak2006five}, including kin or group selection, direct or indirect reciprocity, and network effects among others~\cite{chen2015first,tilman2020evolutionary,weitz2016oscillating,hilbe2018evolution,shao2019evolutionary}. Critically, these do not describe why pro-social behavior originates, but rather how it can spread through a population once it is present. Some studies have shown the emergence of cooperation can emerge without these factors, for example through a combination of minimizing payoff differences between players and maximizing the sum of payoffs \cite{mcavoy2022evolutionary}. Other work has studied the effect of social norms, understood as a sequence of rules for updating status based on actions and the status of those interacting, finding criteria for reputation dynamics that maintain cooperation \cite{ohtsuki2006leading}. Researchers have also considered player's acting to maximize a linear combination of their payoff and that of their opponent, in a spatial game, considering a broad range of symmetric two player two action games \cite{szabo2013coexistence, szabo2012selfishness}. Similar studies have focused on lattice games; however all of these have the potential for spatial assortativity to select for cooperation~\cite{nowak1992evolutionary,szabo1998evolutionary,perc2008social,santos2006evolutionary}. We add to this body of work by considering how adherence to social norms can evolve to promote cooperative behaviors in a well-mixed population, removing potential confounding effects, other factors known to allow cooperative behavior to spread through a population. 


\section{Model and Methods}
\begin{figure}
    \centering
    \includegraphics[width=5.5in]{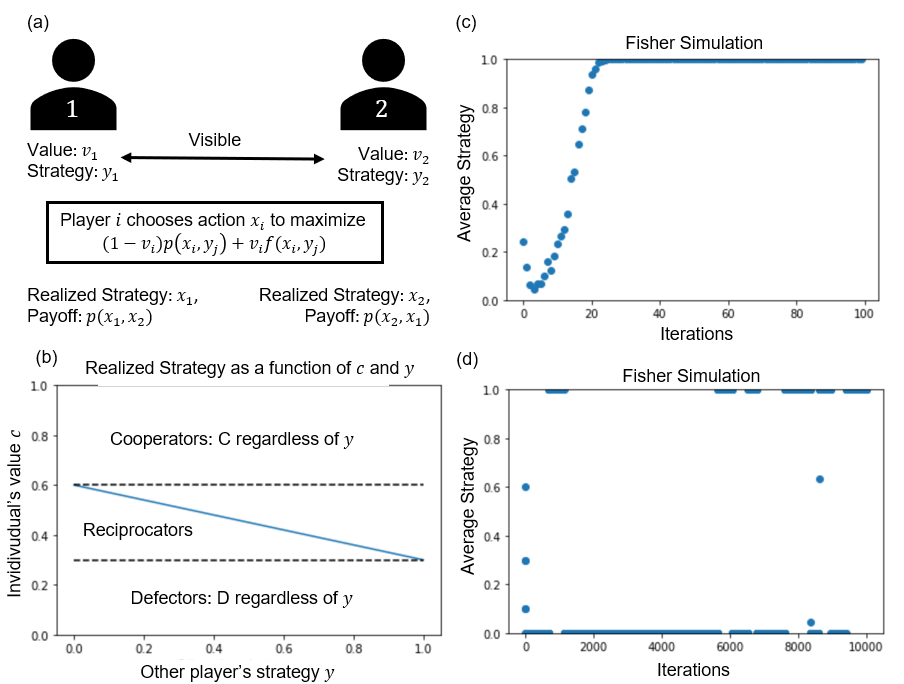}
    \caption{Social norms drive internal deliberation for behavioral responses. (a) we present a diagram of the model. Individuals have a value $v$ and initial strategy $x$. Then when interacting with an individual with strategy $y$, they choose strategy $x$ to maximize their utility $(1-v)p(x,y)+vf(x,y)$, where $p(x,y)$ is the payoff and $f(x,y)$ is some function encoding the norm. Each individual in the pair follows this procedure to determine their realized strategies. We consider mixed strategies $xC+(1-x)D$ in the Prisoner’s Dilemma. (b) we plot the corresponding utility maximizing actions as a function of $y$ and value $c$, used instead of $v$ to differentiate results for each norm. The thresholds $c_0$ and $c_1$ delineate values where players always defect or cooperate from those who match their opponent's strategy. Thus, there are four types of players, the cooperators, defectors, and reciprocators, with $y=0$ or $y=1$. In the two panels on the right, we see a Fisher simulation of this system. (c) shows the short-term dynamics, a population near the threshold transitioning from defection to cooperation. (d) illustrates the long-term behavior of the system under low mutation rates, showing fast transitions between cooperating and defecting states}
    \label{fig-1}
\end{figure}

Individuals have a set value $v$ that determines how important a social norm is to them, measuring either their adherence to this norm, or some notion of niceness described by the norm. Against a player following strategy $y$, an individual with value $v$ chooses the best-response strategy $x$ that maximizes their utility \begin{equation}
    x^* = \textrm{argmin}_x (1-v)p(x,y)+vf(x,y)
\end{equation} where $p(x,y)$ is the payoff to an individual following strategy $x$ with one following strategy $y$, and $f(x,y)$ is a function encoding the social norm, depicted in Fig. \ref{fig-1}. We note that similar introspection dynamics has been studied in prior work~\cite{couto2022introspection}, but our present model focuses on internal deliberations driven by norms. This framework is general enough to encompass many scenarios, including any game, encoded by $p(x,y)$. Depending on the form of $f(x,y)$, many different norms can be represented. Figure \ref{norm-table} gives a list of the four norms we consider here, though many more are possible. Our approach establishes a continuum between those who purely maximize their own individual payoff, $v = 0$, and those who solely follow the social norm, $v=1$, allowing for a smooth transition between these extremes. In addition, the social norm $f(x,y)$ can be seen as a regularization term, which are used broadly in optimization when there is more than one goal, for example models trying to fit data with minimal complexity. 

Note there are several different functions that could encode the same norm. For example, fairness could also be encoded by $f(x) = -|p(x,y)-p(y,x)|$ or the difference to an even power, or some sigmoid functions. One can also include parameters in these, for example a multiple of $x-y$ parameter in the reciprocity norm, which describes the importance of differences in strategy, the harshness of the norm, as the utility drops off faster as strategies become more distant. 

\begin{table}[H]
    \centering
    \begin{tabular}{c|c}
        norm & $f(x,y)$ \\
        \hline
        compassion & $p(y,x)$ \\ 
        univeralizability & $p(x,x)$ \\ 
        reciprocity & $\exp(-(x-y)^2)$ \\ 
        equity (fairness) & $\exp(-(p(x,y)-p(y,x))^2)$ 
    \end{tabular}
    \caption{Different norms can be encoded by various functions $f(x,y)$. Compassion means caring about how one's actions influence others, so individuals care somewhat about maximizing the other player's payoff. Universalizability asks how actions would perform if performed by the whole population, so all will receive payoff $p(x,x)$. Reciprocity means that strategies should be similar, the given function is large when the strategies $x$ and $y$ are closer. Similarly, the equity norm does this with the payoffs of each player, incentivizing these to be close. }
    \label{norm-table}
\end{table}

\vspace{-0.2cm}

The payoff we'll consider comes from the classic Prisoner's Dilemma game, which is the canonical example where cooperation is selected against. We avoid all the features that are known to allow cooperation to emerge, to show this is new, for example spatial structure, to ensure these results are purely an effect of this model of norms. In later plots, we will normalize two parameters, $P = 0$ and $R=1$, of the game so we can plot results in the $ST$-plane. Here, the strategy $y$ is the mixed strategy $yC+(1-y)D$ which cooperates with probability $y$, and otherwise defects. The payoff is then the weighted sum $p(x,y) = Rxy+Sx(1-y)+T(1-x)y+P(1-x)(1-y)$ of possible outcomes, where the matrix in Eq.\ref{pdmatrix} gives the payoff received by the player given their action, in the row, and the other player's action, in the column. 


There are a few important things to note. First, individuals receive the payoff $p(x,y)$, not their utility $u(x,y)$. In this way any nonzero $v$ will usually lead to a decrease in payoff, since only $v=0$ guarantees the payoff maximizing strategy will be chosen. Indeed, increasing $v$ can only decrease payoff, and likely will if $p(x,y)$ and $f(x,y)$ have different maximima. Therefore larger values should theoretically be selected against. Second, individuals cannot perceive the value $v$ of another individual. Consequently, this mechanism cannot be thought of as reputation or green-beard altruism, that is, individuals are somehow cooperating more with "nicer" players, those that also adhere to the norm. 

One consideration in this algorithm is how strategies are updated. It may be unrealistic for strategies to change that quickly, so instead individuals could take a step of some fixed size towards the optimum, or some fixed interpolation between the two (making their strategy a geometric sum of previous optima). Another approach, that we use in some models, is to give individuals a preferred strategy that is fixed. Ultimately, strategies are visible, because in any of these cases, it could be learned by observing the individuals previous actions.  

We analyze these models theoretically, and also perform a series of experiments. These are simulations of a Fisher process on $n$ individuals. That is, each individual has fitness $e^{\beta p}$, where $\beta$ is selection strength, and $p$ is the payoff averaged over all possible interactions in the population. Then $n$ are selected proportional to their fitness for the new population, possibly with some mutation on strategy or trait value. The code for these experiments and the corresponding figures is available at https://github.com/bmDart/CURE. 

We begin by investigating the compassion norm, showing the continuous system has a discrete analog, which we can analyze in two simplifying limits. Then we discuss how this generalizes and the challenges in studying other norms. We will use different variables to clarify which norms each result holds for, for example $c$ for compassion and $u$ for universalizability. 

\vspace{1cm}


\section{Results}
\subsection{Compassion}
In the compassion norm, both $f(x,y) = p(y,x)$ and $p(x,y)$ are linear in $x$, so the utility function is also linear in $x$. Consequently the optimal actions will always be a pure strategy $0$, pure defection, or $1$, pure cooperation (or all actions have equal payoff, when this line has slope zero). A transition between these occurs when their payoffs are equal, when the compassion $c_y$ satisfies $(1-c_y)p(0,y) + c_yp(y,0) = (1-c_y)p(1,y) + c_yp(1,y)$. Solving this yields  \begin{equation}
    c_y = \frac{P-S}{T-S} +  y\frac{\left(T+S-R-P\right)}{T-S} \quad\to\quad c_0 = \frac{P-S}{T-S}, \quad\quad\quad c_1 = \frac{T-R}{T-S}
\end{equation} We can plot this as before to determine the utility maximizing responses, in figure \ref{fig-1}. 


We see there are three distinct regions of the compassion value, For low compassion values below $c_1$, no level of cooperation by the other player will get an individual to cooperate. Similarly, those with high compassion values above $c_0$ will cooperate regardless of the other players strategy. For intermediate levels, the utility maximizing strategy depends on the other player's probability to cooperate. However, as noted earlier, the optimal values are always either pure cooperation or pure defection. For these, all individuals with values in the range $(c_1,c_0)$ will defect against a defector and cooperate with a cooperator. Because of this, we'll call them reciprocating strategies. One could also consider them similar to the Tit-for-Tat strategy in the iterated prisoner's dilemma which copies it's opponent's previous move~\cite{wolf2011coevolution,schmid2021unified}. Thus this system with continuous values and strategies amounts to a discrete system with four kind of players: those who always cooperate or defect, $C$ or $D$, and those who reciprocate and prefer cooperation or defection, $R_C$ or $R_D$. While both types cooperate with $C$ and defect against $D$, the first type of reciprocator will cooperate with itself, whereas the second will not. This system is minimal enough to analyze theoretically, though we find simulation is necessary for norms exhibiting more complicated dynamics. We discuss this discrete model in the next section. 

Intuitively, the reason why this model of norms may be able to promote cooperation is simple. One can imagine a population where everyone is defecting. Values below $c_0$ will be under neutral selection, since they will follow the same strategy, defection, as all other individuals. However, when an individual with a value above $c_0$ emerges, they will cooperate with everyone. Normally, this would be disadvantageous, since defectors will exploit this. However, the reciprocating players will serve as a buffer, cooperating with the cooperators to increase their payoffs and defecting against the defectors to decrease theirs. Depending on the portion of reciprocating players when a cooperator emerges, and the parameters of the game, a wide variety of dynamics can be seen. It turns out not to be sufficient for the cooperator to initially have higher average payoff than a defector, as it's possible the growing number of cooperators will make the population vulnerable to the remaining defectors. In the next two sections, we will analyze this system under two simplifying limits, low selection and large population size.  

Before analyzing this further, we first note that the slope of the dividing line $c_y$ is $\frac{\left(T+S-R-P\right)}{T-S}$, which is negative if and only if \begin{equation}\label{C1}
    T+S < R+P < 2R
\end{equation} However it's possible for this condition to fail even though the game is still a prisoner's dilemma, for example with $(S,P,R,T) = (0, 0.1,0.2, 1)$. In such cases, individuals with intermediate values will cooperate with defectors and defect against cooperators. This counter-intuitive behavior results from the fact that $c_0$ and $c_1$ are essentially the differences $P-S$ and $T-R$, just normalized by $T-S$. These are the amounts gained by defecting against a defector or cooperator, respectively. Since compassion must outweigh the benefit of defection, these differences determine the necessary level of compassion for a player to cooperate. Therefore if it costs more to cooperate with a cooperator than a defector, players will need to be more compassionate to do so. We will not study this case, as such behavior cannot promote cooperation. Rather than acting as a buffer that promotes cooperating individuals, as in the previous case, individuals with intermediate values will promote defectors at a cost to themselves while hurting cooperators, leading to their extinction. Interestingly, this same condition delineates two types of behavior in the universalizability norm, which we discuss later.

\subsection{Monomorphic Populations}
In the limit of weak selection, Antal et al. derived a condition for traits to be favored based on the entries of a matrix of interaction payoffs under arbitrary mutation rates \cite{antal2009mutation}. Specifically, they determine a necessary condition for a specific trait to be present than $1/n$, if there are $n$ straits, in the mutation-selection equilibrium with arbitrary levels of mutation. We can think of the discrete model as one of these games with four types, with interaction payoffs given below. 

\begin{center}
    \begin{tabular}{c|cccc}
         & $C$ & $R_C$ & $R_D$ & $D$ \\
         \hline
         $C$   & $R$ & $R$ & $R$ & $S$\\ 
         $R_C$ & $R$ & $R$ & $T$ & $P$ \\ 
         $R_D$ & $R$ & $S$ & $P$ & $P$ \\ 
         $D$   & $T$ & $P$ & $P$ & $P$ \\ 
    \end{tabular}
\end{center} 

Applying their condition for low mutation rates, we see that $C$ is favored when $S+2R >  T+2P$. In the normalization $P=0$, $R=1$ this corresponds to the region $S > T-2$, a triangle next to the origin. This makes sense, as it means the payoff to a cooperator who is defected against, $S$, must be large, and the payoff to a player who defects against a cooperator, $T$, must be small. The type $D$ is favored in the complementary region. Next, $R_C$ is favored when $T+2R > S+2P$, which is always true, and $R_D$ is never favored. Lastly, $R_C$ will always be more frequent than $D$ in equilibrium. This all holds as well for the high mutation rate condition, though now the condition for $C$ is weaker, $S + 3R > T+3P$, corresponding to a larger set of $R,P$ values. Since the conditions extend linearly to any intermediate value of mutation, this means $R_C$ will always be favored the most, and $C$ can be favored when $D$ is not, depending on the payoffs and mutation level. This makes sense, as they perform well with both cooperators and defectors, and exploit $R_D$.  This model is slightly inaccurate, as the paper assumes uniformly random mutation, which won't come from mutation on the players' values, for example $D$ is more likely to mutate to $R_D$ or $R_C$ than $C$, as this former require less of an increase in their value. 



\begin{figure}
    \centering
    \includegraphics[width=5.5in]{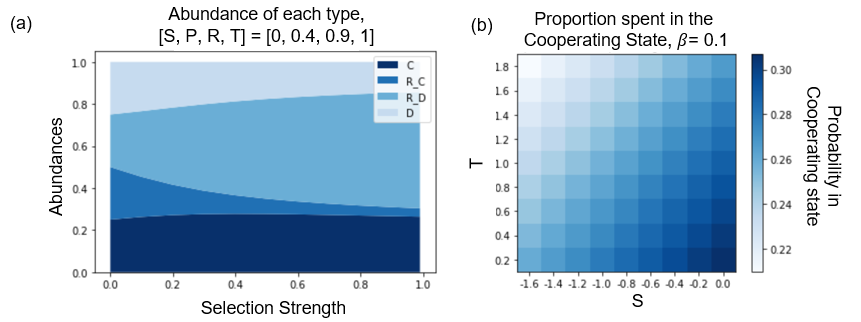} \hspace{1cm} 
    \caption{The optimal selection strength to promote unconditional cooperation. (a) plots the proportion of time the population spends in each state for a particular game and varying levels of selection strength. Surprisingly, there is a slightly non-monotonic relationship, at some point increasing selection causes cooperators to be favored less. Further, increasing selection seems to only result in lower quantities of cooperating players, $C+R_C$. (b) shows a heatplot of the proportion for the cooperating state over all possible games, where $P=0$ and $R=1$ so the space of games is two dimensional, the values of $S$ and $T$. We see that this quantity is effectively determined by $S-T$. As expected, high values for $S$, the payoff of a cooperator when defected against, and low values of $T$, the payoff of a defector when defecting against a cooperator, yields the largest levels of cooperation.}
    \label{fig-2}
\end{figure}

Conversely, we can model in the limit of low mutation but with arbitrary selection~\cite{fudenberg2006imitation}. In this case, the population will be mostly monomorphic, solely consisting of one of the types $C$, $R_C$, $R_D$, and $D$. When a mutant emerges, we can calculate the fixation probability by the classic fixation formulae~\cite{traulsen2006stochastic} \begin{equation}
    \frac{1}{1+\sum_{j=1}^{N-1} \prod_{i=1}^{j} \frac{g_i}{f_i}}
\end{equation} where $f_i$ is the probability of transitioning from state $i$ to state $i+1$, and $g_i$ is the probability of transitioning from state $i$ to $i-1$, where the state is the number of invaders. In particular, these are \begin{align*}
    f_i &= \exp\left(\beta\frac{a(i-1)+b(N-i)}{N-1}\right) \\
    g_i &= \exp\left(\beta\frac{ci+d(N-i-1)}{N-1}\right)
\end{align*} where $\beta$ is the strength of selection, and $a$, $b$ are the payoffs of an invader interacting with an invader or resident, respectively, and $c$, $d$ are those for a resident. Interestingly, using a linear fitness, interpolating $1$ and the the term in the exponent, qualitatively similar results are obtained (the benefit of this approach is that fitness will always be positive). The cases $C \leftrightarrow D$ and $D \rightarrow C$ are classical, while $C \leftrightarrow R_C$ and $D \leftrightarrow R_D$ are neutral drift, as both types always choose the same action. We can consider $R_D \to C$ and $R_C \to D$ as neutral drift, as the invading reciprocator will switch to the resident strategy after the first interaction. The most complex is when a $D$ invades $R_C$, or $C$ invades $R_D$, as the other reciprocating types would be created by interactions. Nonetheless, we can approximate the fitness assuming as the expected value over interactions, for example, $R_C$ will cooperate with themselves and defect against $D$, so their fitness would be a weighted average of these possibilities by the relative frequency of $R_C$ and $D$. Also, $R_C$ invading $D$ is impossible, as emerging reciprocating players always follow the resident strategy, unless mutation occurs simultaneously on value and strategy. Another technical issues is that we cannot have $\beta=1$, since the game parameters would make the expression have a division by zero term. On this timescale, the population is a Markov Chain over the monomorphic states with the transition matrix $$\begin{bmatrix}
    P_{C \to C} & P_{C \to R_C} & P_{C \to R_D} & P_{C \to D} \\
    P_{R_C \to C} & P_{R_C \to R_C} & P_{R_C \to R_D} & P_{R_C \to D} \\
    P_{R_D \to C} & P_{R_D \to R_C} & P_{R_D \to R_D} & P_{R_D \to D} \\ 
    P_{D \to C} & P_{D \to R_C} & P_{D \to R_D} & P_{D \to D}
\end{bmatrix}$$ We can then look at it's principal eigenvector to see the proportion of time the chain spends in each state, to get a sense of which are favored. This will vary with selection strength $\beta$ and the game parameters $P$ and $R$, so to get a sensible plot, we'll take slices of the parameter space. This  is done in figure \ref{fig-2}, where first the selection strength $\beta$ is varied for a fixed game, then the key game parameters $S$ and $T$ are varied for a fixed selection strength. 


\subsection{Large Population limit}
\begin{figure}
    \centering
    \includegraphics[width=5.5in]{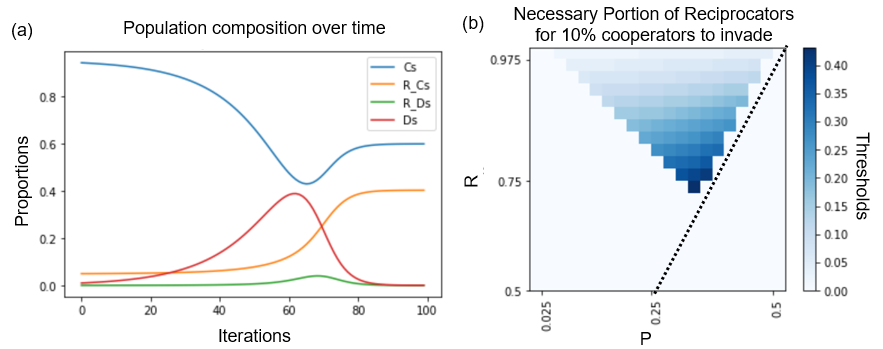}
    \caption{Deterministic dynamics under large population limit. (a) Here we see a model under an infinitely large population, to remove stochasticity. There are rich dynamics between the various type. In (a) we plot the proportions of each type over time, starting from a cooperating population with a small number of reciprocating players, invaded by a small number of defectors. The cooperators initially have lower payoff, due to the small number of reciprocators, and decline. Defectors initially are taking over, but this creates a sufficient number of reciprocators to counteract this, around $t=60$ iterations. The defectors then have lower payoff, allowing cooperators to recover, and the invading defectors to go extinct. Crucially, if there were too few reciprocators, the defectors would be able to fixate and replace the cooperators; this plot demonstrates the behaviour around this threshold. This threshold will vary for each game. (b) we plot this threshold for every game, normalized to have two parameters. As expected, higher $R$, reward for mutual cooperation, helps cooperators, so less reciprocators are necessary for them to invade a defecting population. Interestingly, there is less effect of $P$, the punishment payoff for mutual defection. Additionally, we see a barrier indicating some games where cooperators cannot invade, at least in this model. In addition, we see a line $R=2P$ that separates cases where this level of cooperators can invade in this model. Thus, there are some games where no amount of reciprocating players is enough for cooperators to invade. }
    \label{fig-3}
\end{figure}

In an infinitely large well-mixed population, the number of interacting pairs is proportional to the product of the frequencies of each individual in the pair. For example, of the $C$ cooperators, $C/N$ would interact with other cooperators, getting a payoff of $R$, $R_C/N$ would interact with reciprocating cooperators, getting a payoff of $R$, and so on. This gives each type their average fitness. The new proportions will be (before normalization) \begin{align} 
    C'   &= C(R(C+R_C+R_D)+SD) \\
    R_C' &= R((R_C+R_D)C+R_CR_C)+SR_DR_C \\
    R_D' &= P((R_C+R_D)D+R_DR_D)+TR_CR_D \\
    D'   &= D(TC+P(R_C+R_D+D))
\end{align} The first variable of each term is the focal player and the second is the other player. For example, $C$ will get the payoff $R$ when interacting with a $C$, $R_C$, or $R_D$ player, giving the first three terms of the first equation. The second equation indicates that an $R_C$ or $R_D$ player interacting with a $C$, or two $R_C$'s together, get a payoff of $R$ and become and $R_C$, and the last terms says that an $R_D$ will encounter a $R_C$, become and $R_C$ and attempt to cooperate, but receive a payoff of $S$, as the other player saw and $R_D$ and so defected. The next two equation can be interpreted similarly. This systems does not appear to have an analytic solution, and exhibits complicated behavior, see figure \ref{fig-3}. This demonstrates that an initially higher payoff for cooperators need not guarantee they will invade successfully. Then, by comparing the final proportions of each type for various initial proportion of reciprocators, one can determine what amount of reciprocators is necessary to allow cooperators to invade a defecting population, or defectors to invade a cooperating population, in figure \ref{fig-3}. 

\subsection{Other norms}
The universalizability norm is similar. Now, the utility function is $(1-u)p(x,y)+up(x,x)$. Note that this is quadratic in $x$, so if it is concave up, the maximal values will be achieved at the endpoints $x=0$ or $x=1$. This means the same approach as in the compassion case may be used. In particular, the threshold level of universalizability $u_y$ against a player with strategy $y$ makes the utility equal at the endpoints, giving \begin{equation}
    u_y = \frac{P-S-y(R-S-T+P)}{R-S-y(R-S-T+P)}, \quad\quad u_0 = \frac{P-S}{R-S}, \quad\quad u_1 = \frac{R-T}{P-T}
\end{equation} This is part of a hyperbola, and creates the same three regions as in the compassion case. Thus, the same system of cooperators, reciprocators, and defectors may be used, the only difference being the locations of the thresholds as a function of the game parameters. Comparing these thresholds with the compassion case, we see that $u_1 < c_1$ and $u_0 > c_0$, which means that for a set game, lower values are required to cooperate with a cooperator, and higher values are needed to cooperate with a defector. Thus, individuals in this norm are quicker to help cooperators and punish defectors. Equivalently, the region of values corresponding to reciprocating players is larger, thus there is a stronger buffer effect, promoting cooperation further (at least in the region $R+P > S+T$). Indeed, simulations find that cooperation can invade a defecting population in a broader range of circumstances than in the compassion norm. This is consistent with the discussion in the introduction noting that universalizability was a stricter norm than compassion. 

The concave down case is more complicated, as now intermediate strategies maximize player's utility. This occurs when the coefficient of $x^2$ in the utility function, $R-S-T+P$, is negative. Interestingly, this is the opposite of condition \ref{C1}, which delineated cases in the compassion norm. Intermediate levels of cooperation could result in a continuous transition towards cooperation. Now, the utility maximizing action as a function of $u$ and $y$ goes from $0$ at $u=0$ to the maximum of $p(x,x)$, which depends on the game's parameters. This is actually interesting in and of itself, as it turns out that cooperation need not be the optimal solution for a population in the prisoner's dilemma. The conclusion of the classical prisoner's Dilemma is that while defection is optimal for an individual player, it is better for the pair to mutually cooperate. It is surprising, then, when this fails to hold for the larger population. For some parameter choices, more can be gained by the exploitative defector-cooperator interactions than is lost by the mutual defection interactions (indeed, it is possible for any level of cooperation to be optimal for a population). Interestingly, the optimal value $-\frac{S+T-2P}{2(R-S-T+P)}$ of $p(x,x)$ is similar to the fixed point $\frac{P-S}{R-S-T+P}$ of the replicator dynamics of this game. Simulation shows cooperation can not emerge, even in the most favorable circumstances, but it can be maintained in some cases. While not as clearly impossible as in the compassion case, this region still suppresses the evolution of cooperation. 

The reciprocity norm has utility $(1-r)p(x,y)+re^{-(x-y)^2}$. Similar to the concave down case in universalizability, the utility maximizing strategy increases from zero at $r=0$ to $y$ at $u=1$, with speeds depending on the game parameters, since $x=y$ maximizes $f(x,y)$. Thus, we study this case by simulation, finding cooperation would never invade, and indeed defection would always fixate in a cooperating population. Intuitively, high values of this norm promote taking a similar strategy as your opponent, which means defectors will be defected against and cooperated will be cooperated with. However, there is no mechanism to introduce cooperation into a defecting population. 

Lastly, the equity norm turns out to just be a sharper version of the reciprocity norm, for the given functions. Indeed, the norm can be simplified to $f(x,y) = e^{-(S-T)^2(x-y)^2}$, since $p(x,y)-p(y,x) = Sx(1-y)+T(1-x)y-(Sy(1-x)+T(1-y)x) = (S-T)[x(1-y)-y(1-x)] = (S-T)(x-y)$, canceling some terms. This is interesting in and of itself, as it shows a connection between seemingly different notions of morality. Because of this connection, equity should behave qualitatively similar to the reciprocity norm. Studying a more sensitive version of a previous norm gives initial results investigating parameterized norms and the effects of changing parameters. However, since reciprocity showed no cooperation, the same was found here. 

\begin{figure}
    \centering
    \includegraphics[width=5in]{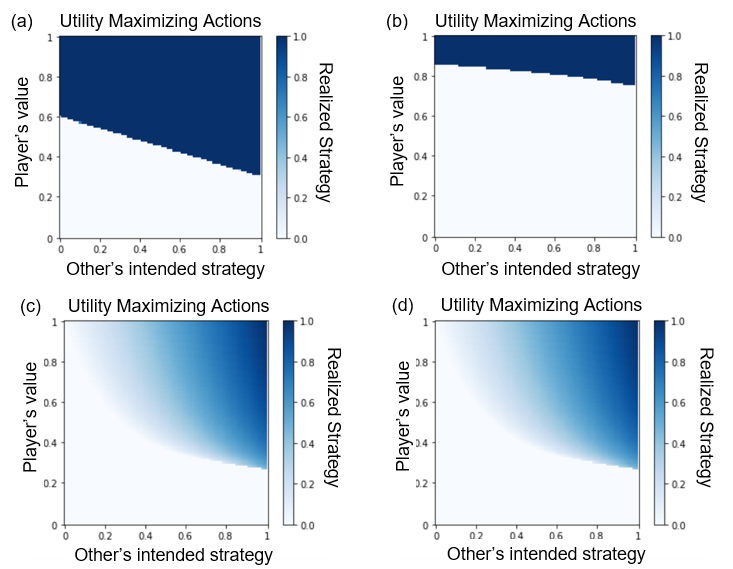}
    \caption{Strategy dynamics in all four norms. This plot is a comparison of the norms for the game S=0, P=0.6, R=0.7, T=1. (a) and (b) we see the first two usual have realized strategies that are pure, either zero or one, and an agreement with the earlier results. (c) and (d) result in mixed strategies. Here they are the same, because our choice of $f(x,y)$ makes one essentially a multiple by $S-T$ of the other, which is one for this game, so they are the same.}
    \label{fig-4}
\end{figure}

A good way to compare these norms is by plotting the best responses in each, for a fixed game, see figure \ref{fig-4}. As in \ref{fig-1}, the utility maximizing action is plotted as a function of the other player's intended strategy, on the horizontal axis, and their value, on the vertical axis. 

\newpage

\section{Discussion and Conclusion}
Much work has been done throughout philosophy, psychology, and sociology to understand how norms shape our behavior and evolve over time~\cite{gachter2009reciprocity}. Our study complements this qualitative treatment with a mathematical formalism that extends previous work on the origin of cooperative behavior. We propose a model where individuals choose actions depending on the other player's strategy and a value describing their adherence to some cultural norm, to maximize a combination of their own payoff and a expression encoding the norm. This captures the fact that decisions are often made by considering more than just the literal payoff of one's actions. We then investigate this model from numerous angles. Using simplifying limits, we are able to obtain analytical results in a discrete description to our model in finite populations. Alternatively, we also consider infinitely large populations, where the model becomes deterministic. This is analyzed numerically to determine the necessary number proportion of our reciprocating players to allow for cooperation to spread in a population.    

This work introduces a framework that encompasses decision making under a wide variety of social norms, by encoding these in a general function. We investigate whether this mechanism can promote cooperation without the presence of other factors known to do this, such as spatial structure or other forms of assortment. Further, this approach also allows us to analyze different norms from the same perspective, to compare then and even consider their evolution. As such, future work can investigate profound questions like which norms can replace others, how do norms change over time, and is there a best norm for promoting cooperation? Our framework also allows for a continuous emergence of cooperation, as compared to earlier work often considers discrete traits. 

In the compassion norm, we see cooperation selected for in a large number of versions of the Prisoner's Dilemma, given by different game parameters. This shows how our proposed mechanism can resolve the dilemma in many varied cases~\cite{leimar2001evolution,schmid2021unified,shu2023determinants}, though does not guarantee the emergence of pro-social behavior. The essential reason we see this is that the reciprocating players serve as a buffer in the population. They simultaneously defect against defectors, and cooperate with cooperators, suppressing the former while boosting the latter. A stronger effect is seen in the universalizability norm, due to its similarity. However, the reciprocity and equity norms are unable to promote pro-social behavior, because they provide no incentive to be more cooperative than the other player. Comparing this diverse set of norms, we establish the possibility, but not guarantee, for norms to allow for the emergence of pro-social behavior. %

Our work presents a mathematical model of decision making under general social norms, where individuals compare the payoff they receive to how closely they achieve the orders of their norm. Quantifying this by a parameter allows for a continuum between purely selfish and purely selfless individuals, in contrast to previous models which often assume distinct behaviors between these two groups. Using this framework, we investigate whether cooperation can emerge and spread through a population depending on the social norm and game being played. 

First studying the compassion norm, we note this continuous model reduces to a discrete system. This can be analyzed in the simplifying limit of weak selection, where we find a condition for the cooperators to be favored. Alternatively, under the limit of low mutation and arbitrary selection, the population is monomorphic most of the time, so we can calculate the transition probabilities between states to determine the proportion of time the population stays in each state. We find that this is essentially determined by a simple combination of the game parameters. Then we consider a large population limit, where all interactions become averaged to remove stochastic effects. Doing so, we are able to compute the levels of reciprocating players necessary for a small number of cooperators to fixate in a population. This increases as the game becomes less favorable to cooperators, until eventually no amount of reciprocating players are sufficient. 

We then consider how this analysis may be extended to other norms. Universalizability is a similar norm, in that a discrete systems emerges in this case as well, though the thresholds for this categorization are further apart than in the compassion norm. Consequently, the ability to promote cooperation in these cases is even greater. However, an alternative regime occurs where the optimal response varies continuously with adherence to the norm, preventing a discrete model of the system. Simulation in these cases found cooperation could not emerge. In the reciprocity norm, the actions chosen vary from defection to matching the other players strategy, as the player's value increases from zero to one. Simulation shows no cooperation is able to emerge, as there is only an incentive to cooperate as much as the other player, but no further. As a result, the players who cooperate less on average do better overall, and the level of cooperation falls over time. The same holds in the equity norm, as some straightforward algebra shows it is essentially a stronger version of the reciprocity norm, at least using the form we study. Thus, the simulation results are similar. By studying a range of different social norms grounded in the psychological tradition, we demonstrate a widely varying ability to explain the emergence of cooperation depending on the norm and game under consideration. 

Future work in this framework could investigate a number of interesting questions. Thanks to our model's generality, it can be applied to practically any game, for example, the ultimatum~\cite{page2000spatial,wu2013adaptive} and public goods games~\cite{kurokawa2009emergence} are to other widely studied examples used to understand strategic behavior in different contexts, for example, via multilayer interactions~\cite{zhang2015cooperation,su2022evolution}. Initial simulation results suggest that the compassion norm can allow agents to find the optimal coordinate equilibrium, a result which may extend to other coordination games. Additionally, there are many functions that could encode a given social norms, and many other high-order social norms that could be considered~\cite{leimar2001evolution,santos2018social,kessinger2023evolution}. The precise form of our results depend on the particulars of this function, but perhaps general insights can be gained by thinking of what forms these functions might take, such as which features are necessary to promote cooperation. Perhaps one could even evolve the norm of a function itself to see which are optimal for certain games. By putting different norms, compassion, universalizability, reciprocity, and equity, in the same framework, one can see which are better able to promote cooperation. One could even combine norms by incorporating additional values for each player governing how much they follow each norm. That way, even if one norm may be unable to promote cooperation in isolation, it may be able to in conjunction with another. 

The evolution of pro-social behavior has been a longstanding question in biology with numerous explanations proposed through the years~\cite{gachter2009reciprocity,santos2021complexity}. By considering a well mixed population with minimal other factors, we have demonstrated the bottom-up emergence of cooperation under the influence of top-down social norms. In light of growing concerns regarding potent AI systems and their impact on humanity~\cite{dafoe2021cooperative}, our work paves the initial way for leveraging built-in behavioral norms to moderate the cooperation of artificial agents~\cite{mcnally2012cooperation} in hybrid AI-human systems~\cite{barfuss2020caring,leonard2022collective,traulsen2023future,chen2023ensuring}.   




\printbibliography

\end{document}